\documentclass[aps,showpacs,prc,twocolumn]{revtex4}

\usepackage{graphicx}
\usepackage{amssymb}
\newcommand{\ba}{\begin{eqnarray}}
\newcommand{\ea}{\end{eqnarray}}
\newcommand{\bmath}{\begin{mathletters}}
\newcommand{\emath}{\end{mathletters}}
\newcommand{\ban}{\begin{eqnarray*}}
\newcommand{\ean}{\end{eqnarray*}}

\begin{document}

\title{First-Order Quantum Phase Transition in a Finite System}

\author{A. Leviatan}

\affiliation{
Racah Institute of Physics, The Hebrew University,
Jerusalem 91904, Israel}

\date{\today}

\begin{abstract}
The dynamics at the critical-point of a 
general first-order quantum phase transition in a finite system 
is examined, from an algebraic perspective. Suitable Hamiltonians 
are constructed whose spectra exhibit coexistence of states corresponding 
to two degenerate minima in the energy surface separated by an arbitrary 
barrier. Explicit expressions are derived for wave 
functions and observables at the critical-point. 
\end{abstract}


\maketitle

\newpage

Quantum phase transitions 
have become a topic of great interest in different 
branches of physics~\cite{iac00,iac01,voj03}.
These are structural changes occurring at zero temperature as a 
function of a coupling constant and driven by quantum fluctuations. 
The phase transitions are said to be of second-order when the 
system changes continuously from one phase to another, and 
of first-order when there is 
a discontinuous change between two coexisting phases. 
Advanced experiments have identified such structural changes 
in a variety of mesoscopic systems, 
{\it e.g.}, nuclei, molecules and atomic clusters, where the transitions 
are between different shapes or geometric configurations.  
An important issue in these systems 
is to understand the modifications at criticality due to 
their finite number of constituents. In the present work we 
study this question in connection with nuclei exemplifying a finite 
system undergoing a general first-order shape-phase transition. 

The role of a finite number of nucleons can be addressed in the 
interacting boson model (IBM)~\cite{ibm}, which describes low-lying 
quadrupole collective states in nuclei in terms of a system of $N$ 
monopole ($s$) and quadrupole ($d$) bosons representing valence nucleon 
pairs. 
The model is based on a U(6) spectrum generating algebra and its 
three solvable dynamical symmetry limits: $U(5)$, $SU(3)$, and $O(6)$, 
describe the dynamics of stable nuclear shapes: spherical, axially-deformed, 
and $\gamma$-unstable deformed.
A geometric visualization of the model 
is obtained by 
an energy surface defined by 
the expectation value of the Hamiltonian in the coherent (intrinsic) 
state~\cite{gino80,diep80}
\ba
\vert\,\beta,\gamma ; N \rangle &=&
(N!)^{-1/2}(b^{\dagger}_{c})^N\,\vert 0\,\rangle ~,
\label{cond}
\ea
where $b^{\dagger}_{c} = (1+\beta^2)^{-1/2}[\beta\cos\gamma 
d^{\dagger}_{0} + \beta\sin{\gamma} 
( d^{\dagger}_{2} + d^{\dagger}_{-2})/\sqrt{2} + s^{\dagger}]$.
For the general IBM Hamiltonian with one- and two-body interactions, the 
energy surface takes the form
\ba
E(\beta,\gamma) = E_0 + 
N(N-1)
\frac{\left [ a\beta^{2} - b\beta^3\cos 3\gamma + c\beta^4\right ]}
{(1+\beta^2)^2} ~.
\,
\label{eint}
\ea
The coefficients $E_0,a,b,c$ involve particular linear 
combinations of the Hamiltonian's parameters~\cite{kirlev85,lev87}. 
The quadrupole shape parameters in the 
intrinsic state characterize the associated equilibrium shape. 
For $\beta>0$ the intrinsic state is deformed and represents, 
in a variational sense~\cite{hagino03}, a ground band whose rotational 
members are obtained by standard angular momentum projection.
Particularly relevant for the present work are states of 
good $O(3)$ symmetry $L$ projected 
from the prolate-deformed intrinsic state 
$\vert \beta,\gamma=0;N\rangle$ of Eq.~(\ref{cond}), 
\ba
\vert\, \beta; N, L,M\rangle &\propto&
\left [\Gamma_{N}^{(L)}(\beta)\right ]^{-1/2} 
\hat{\cal{P}}_{LM}\vert \beta,\gamma=0; N\rangle
\nonumber\\
\Gamma_{N}^{(L)}(\beta) &=& 
\frac{1}{N!}\int_{0}^{1}dx 
\left [ 1 + \beta^2\,P_{2}(x)\right ]^N P_{L}(x) ~. \qquad
\label{wfqpt1}
\ea
Here $P_{L}(x)$ is a Legendre polynomial with $L$ even 
($0\leq L\leq 2N$) 
and $\Gamma_{N}^{(L)}(\beta)$ is a normalization factor.
The projected states with fixed $N$ and $L$, involve a mixture 
of components $\vert N,n_d,\tau,L\rangle$ with quantum numbers related 
to the $U(6)\supset U(5)\supset O(5)\supset O(3)$ chain.
In general the $L$-projected states $\vert\, \beta; N, L\rangle$ 
interpolate between the $U(5)$ spherical ground state, 
$\vert s^N\rangle\equiv \vert N,n_d=\tau=L=0\rangle$, 
at $\beta=0$, and the prolate-deformed ground band 
with $SU(3)$ character 
$(\lambda,\mu)=(2N,0)$, at $\beta=\sqrt{2}$. 

Shape-phase transitions in nuclei have been studied 
in the geometric framework of a Bohr Hamiltonian for macroscopic 
quadrupole shapes. Analytic solutions, based on infinite 
square-well potentials,
called E(5)~\cite{iac00} and X(5)~\cite{iac01}, 
have been shown to be relevant to the dynamics at 
the critical-point of second- and first- order transitions respectively, 
and empirical examples have been presented~\cite{caszam0001}. 
Phase transitions for finite N can be studied 
by an IBM Hamiltonian of the form, $H_1 + gH_2$, 
involving terms from different dynamical symmetry chains~\cite{diep80}. 
The nature of the phase transition is 
governed by the topology of the corresponding 
energy surface (\ref{eint}), which serves as a Landau's potential. 
IBM Hamiltonians of this type have been studied 
extensively~\cite{izc98,ckz99,zam02,levgin03,arias03,
iaczam04,rowe04,dusuel05,lev05}, 
revealing the finite-N attributes of the phase transitions. 
The critical $U(5)$-$O(6)$ Hamiltonian was found to exhibit 
a $\gamma$-independent flat-bottomed energy surface, 
characteristic of a second-order phase transition. The critical 
$U(5)$-$SU(3)$ Hamiltonian displays an energy surface with 
degenerate minima separated by an extremely low barrier, hence 
corresponds to a very specific (and non-generic) first-order phase 
transition.  
\begin{figure}  
\begin{center}
\rotatebox{270}{\includegraphics[scale=0.35]{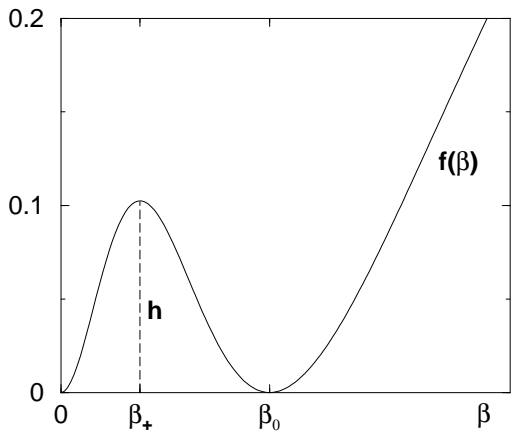}}\hspace{1cm}
\end{center}
\vspace{-0.2cm}
\caption{The energy surface, Eq.~(\ref{ecri1st}), 
at the critical-point of a 
first-order phase transition. 
The position and height of the barrier are 
$\beta_{+} = (-1 + \sqrt{1+\beta_{0}^2}\;)/\beta_0$ and  
$h = f(\beta_{+}) = 
( -1 + \sqrt{1+\beta_{0}^2}\,)^2/4$, with 
$f(0)=f(\beta_0)=0$, $f(\infty)=1$. 
\label{fig1}}
\end{figure}
It is clearly desirable to explore, from an algebraic perspective, 
the finite-N properties 
of a general first-order critical-point with an arbitrary barrier. 
This is the goal of the present investigation.

In a first-order phase transition the Landau's potential has two 
coexisting minima which become degenerate at the critical-point. 
Such a structure involving spherical and prolate-deformed minima 
can be accommodated in the 
energy surface of Eq.~(\ref{eint}) provided 
$a, b>0$ and $b^2=4ac$ at the critical-point. For $\gamma=0$
the corresponding critical energy surface can be transcribed 
in the form
\ba
E_{cri}(\beta,\gamma=0) &=&
E_0 + c\,N(N-1)f(\beta)
\nonumber\\
f(\beta) &=& \beta^2\,(1+\beta^2)^{-2}\,(\beta - \beta_0)^2 ~.
\label{ecri1st}
\ea
$E_{cri}(\beta,\gamma=0)$, shown in Fig.~1, 
behaves quadratically near the minima 
at $\beta=0$ and $\beta=\beta_0 = 2a/b >0$, 
and approaches a constant for large $\beta$. 
The value of $\beta_0$ determines the position $(\beta=\beta_{+})$ 
and height $(h)$ of the barrier separating the two minima in a 
manner given in the caption. 
$h$ and $\beta_{+}$ increase as a function of 
$\beta_0$, and hence both 
low and high barriers can be considered. 
To construct a critical Hamiltonian with such an energy surface, 
it is convenient to resolve it into intrinsic and 
collective parts~\cite{kirlev85,lev87},
\ba
H_{cri} = H_{int} + H_c ~.
\label{resol}
\ea
Such decomposition can be done exactly for any IBM Hamiltonian,  
as elaborated in great detail in ~\cite{kirlev85,lev87}. 
Adapting this procedure to the critical Hamiltonian, 
the intrinsic part ($H_{int}$) is defined to have the equilibrium 
condensate $\vert \,\beta=\beta_0,\gamma=0 ; N\rangle$, Eq.~(\ref{cond}), 
as an exact zero-energy eigenstate and to have an energy surface with 
a structure as in Eq.~(\ref{ecri1st}). 
$H_{int}$ is found to have the form
\ba
H_{int} &=& h_{2}\, 
P^{\dagger}_{2}(\beta_0)\cdot\tilde{P}_{2}(\beta_0) ~,
\label{hint}
\ea
with
$P^{\dagger}_{2\mu}(\beta_0)= 
\beta_{0}\,s^{\dagger}d^{\dagger}_{\mu} + 
\sqrt{7/2}\,\left( d^{\dagger} d^{\dagger}\right )^{(2)}_{\mu}$, 
$\tilde{P}_{2\mu}(\beta_0)=(-1)^{\mu}P_{2,-\mu}(\beta_0)$. 
Its energy surface coefficients of Eq.~(\ref{eint}) are $a=h_2\beta_{0}^2,\, 
b=2h_{2}\beta_{0},\, c=h_2,\,E_0=0$, hence satisfy, for $h_2>0$, the 
aforementioned conditions of a first-order critical-point.
Since $H_{int}$ is rotational-scalar, it has, by construction, 
the $L$-projected states 
$\vert \beta=\beta_0;N,L\rangle$ of Eq.~(\ref{wfqpt1}) 
as solvable deformed eigenstates with energy $E=0$. 
It has also solvable spherical eigenstates:  
$\vert N,n_d=\tau=L=0 \rangle\equiv \vert s^N\rangle$ and 
$\vert N,n_d=\tau=L=3 \rangle$ with energy $E=0$ 
and $E = 3 h_2\left [\beta_{0}^2 (N-3) + 5 \right ]$ respectively.
For large $N$ the spectrum of $H_{int}$ is harmonic, involving 
quadrupole vibrations about the spherical minimum, 
and both $\beta$ and $\gamma$ vibrations about the deformed 
minimum with frequencies $\epsilon$, $\epsilon_{\beta}$ and 
$\epsilon_{\gamma}$ given by~\cite{lev87}
\ba
\epsilon=\epsilon_{\beta}=h_{2}\,\beta_{0}^2 N \;\; , \;\; 
\epsilon_{\gamma} = 9(1+\beta_{0}^2)^{-1}\,\epsilon_{\beta} ~.
\ea 
For the acceptable range $0\leq\beta_0\leq 1.4$, the $\gamma$-band is 
expected to be considerably higher than the $\beta$-band. 
All these features are present in the exact spectrum of $H_{int}$ shown 
in Fig.~2, which 
displays a zero-energy deformed ($K=0$) ground band, degenerate with a 
spherical $(n_d=0)$ ground state. The remaining states are either 
predominantly spherical, or deformed states arranged in several excited 
$K=0$ bands below the $\gamma$ band. The coexistence of spherical and 
deformed states is evident in the right portion of Fig.~2, which shows 
the $n_d$ decomposition of wave functions of selected eigenstates of 
$H_{int}$. The ``deformed'' states show a broad $n_d$ distribution typical 
of a deformed rotor structure. The ``spherical'' states show the 
characteristic dominance of single $n_d$ components that one would expect 
for a spherical vibrator. 
\begin{figure*}  
\begin{center}
\rotatebox{270}{\includegraphics[scale=0.41]{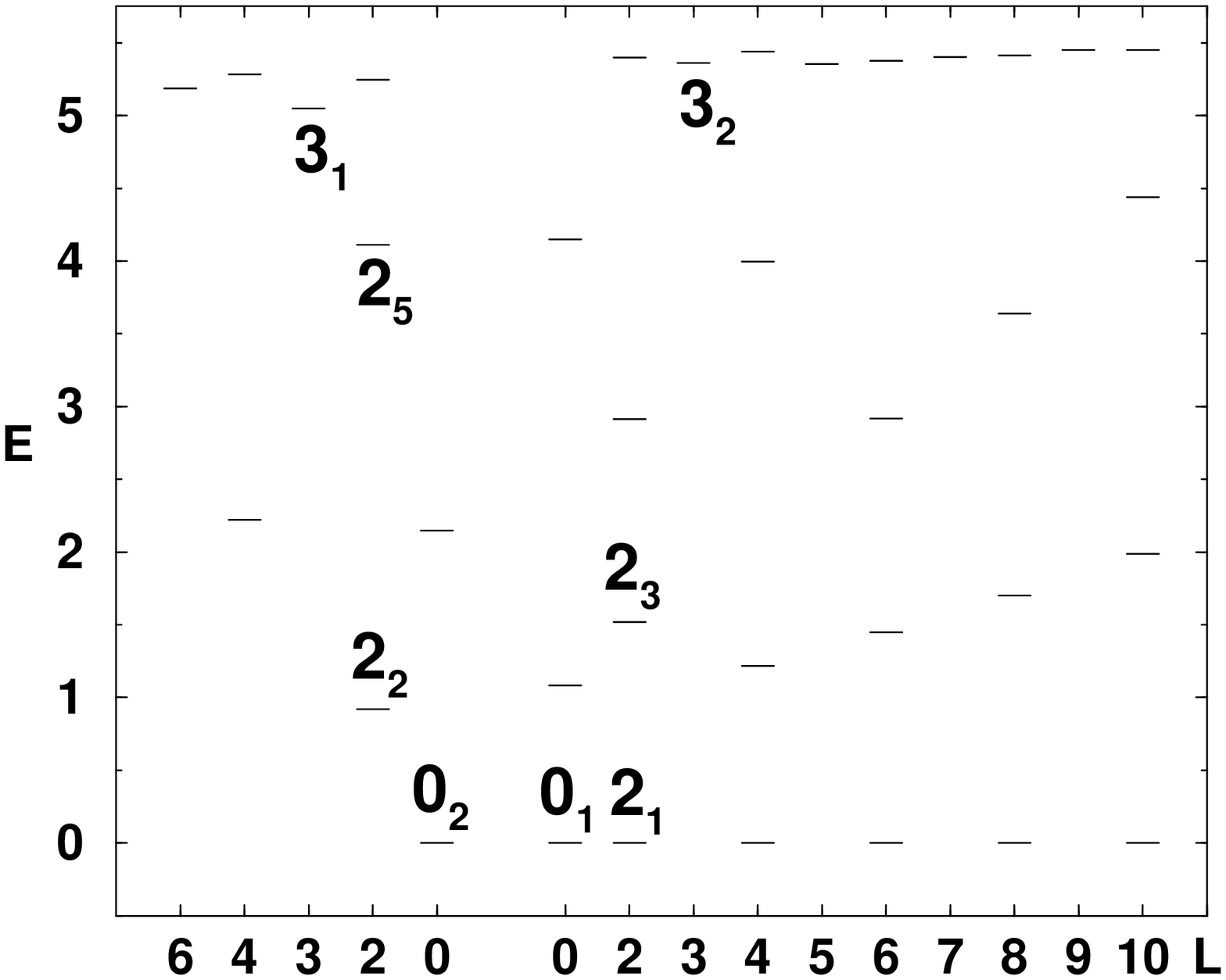}}\hspace{0.4cm}
\rotatebox{270}{\includegraphics[scale=0.41]{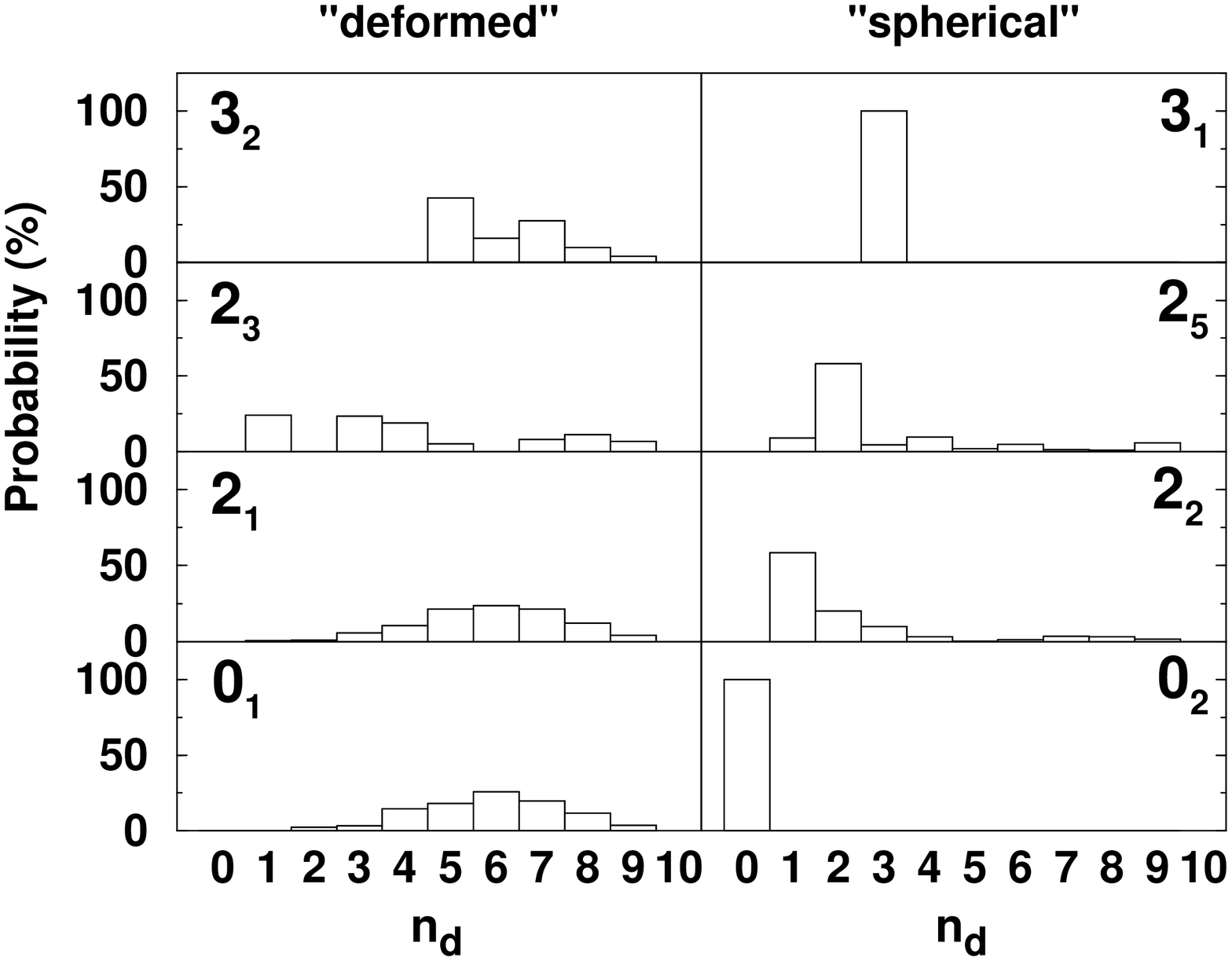}}
\end{center}
\vspace{-0.5cm}
\caption{Left portion: spectrum of $H_{int}$, Eq.~(\ref{hint}), 
with $h_2=0.1$, $\beta_0 =1.3$ and $N=10$. 
Right~portion: the number of $d$ bosons ($n_d$) 
probability distribution for selected eigenstates of $H_{int}$.
\label{f41}}
\end{figure*}

The collective part ($H_c$) 
is composed of kinetic terms which do not affect the shape of the 
energy surface. It can be transcribed in the form~\cite{kirlev85,lev87}
\ba
H_{c} &=& c_3 \left [\, \hat{C}_{O(3)} - 6\hat{n}_d \,\right ]
+ c_5 \left [\, \hat{C}_{O(5)} - 4\hat{n}_d \,\right ]
\nonumber\\ 
&& +\, c_6 \left [\, \hat{C}_{\overline{O(6)}} - 5\hat{N}\,\right ] +E_0~,
\label{hcol}
\ea
where $\hat{N}=\hat{n}_d+\hat{n}_s$, $\hat{n}_d$ and $\hat{n}_s$ are
the total-boson, $d$-boson and $s$-boson number operators respectively.
Here $\hat{C}_{G}$ denotes the quadratic Casimir operator of the 
group G as defined in \cite{lev87}. 
In general, the collective and intrinsic parts of the critical 
Hamiltonian~(\ref{resol}) do not commute. 
Table~I shows the effect of different rotational terms in $H_c$. 
For the high-barrier case considered here, ($\beta_0=1.3$, $h=0.1$), 
the calculated spectrum resembles a rigid-rotor $(E\sim a_{N}L(L+1)$) for 
the $c_3$-term, a rotor with centrifugal stretching 
$(E\sim a_{N}L(L+1) - b_{N}[L(L+1)]^2)$ for the $c_5$-term, and a 
X(5)-like spectrum for the $c_6$-term. In all cases the B(E2) values are 
close to the rigid-rotor Alaga values. This behaviour is different from 
that encountered when the barrier is low, 
{\it e.g.}, for the 
critical $U(5)$-$SU(3)$ Hamiltonian (corresponding to 
$\beta_0=1/2\sqrt{2}$ and $h\approx 10^{-3}$),  
where both the spectrum and E2 rates are similar to the X(5) 
predictions \cite{lev05}. 

To gain more insight of the underlying structure at the critical-point, 
we focus the discussion to the lowest coexisting spherical and deformed 
states. The need for angular momentum projection from mean field solutions 
in regions of shape-coexistence is well known, {\it e.g.}, in 
microscopic calculations employing 
Hartree-Fock Bogoliubov states~\cite{guz00}. In the present case, 
the matrix elements of the critical Hamiltonian (\ref{resol}) 
in the states~(\ref{wfqpt1}),  
$E^{(N)}_{L}(\beta) = \langle\beta;N,L\vert H_{cri}\vert\beta;N,L\rangle
=\tilde{E}_{L}^{(N)}(\beta)+E_0$, 
can be evaluated in closed form
\ba
\tilde{E}_{L}^{(N)}(\beta) &=& 
h_2\,(\beta-\beta_0)^2\,\Sigma_{2,L}^{(N)} + 
c_3\left [ L(L+1) - 6D_{1,L}^{(N)}\right ]
\nonumber\\
&&
+\, c_5\left [ D_{2,L}^{(N)} - \beta^4\,S_{2,L}^{(N)}\right ]
\nonumber\\
&&
+\, c_6\left [ N(N-1) -(1+\beta^2)^2\,S_{2,L}^{(N)}\right ] ~.
\label{eneL}
\ea
Here $D_{1,L}^{(N)}$, $S_{2,L}^{(N)}$, $D_{2,L}^{(N)}$ and 
$\Sigma_{2,L}^{(N)}$ denote the expectation values in 
the states $\vert \beta;N,L\rangle$ 
of $\hat{n}_d$, $\hat{n}_s(\hat{n}_s-1)$, $\hat{n}_d(\hat{n}_d-1)$ 
and $\hat{n}_s\hat{n}_d$ 
respectively. All these quantities are expressed in terms of 
the expectation value of $\hat{n}_s$, denoted by $S^{(N)}_{1,L}$. 
Specifically, 
$D_{1,L}^{(N)}= N - S_{1,L}^{(N)}$, 
$S^{(N)}_{2,L} = S^{(N)}_{1,L}S^{(N-1)}_{1,L}$, 
$\Sigma^{(N)}_{2,L} = (N-1)S^{(N)}_{1,L} - S^{(N)}_{2,L}$, 
$D_{2,L}^{(N)} = N(N-1) - 2(N-1)S^{(N)}_{1,L} + S^{(N)}_{2,L}$.
The quantity $S^{(N)}_{1,L}$ itself is determined by the 
normalization factors of Eq.~(\ref{wfqpt1}) 
\ba
S^{(N)}_{1,L} = \Gamma^{(L)}_{N-1}(\beta)/\Gamma^{(L)}_{N}(\beta) ~.
\ea 
It also satisfies the following recursion relation
{\small
\ba
S^{(N)}_{1,L}
 = \frac{(N-L/2)(2N+L+1)}
{(\beta^4+4)(N-1)+3 
+ (\beta^2-2)(1+\beta^2)S^{(N-1)}_{1,L}}\quad\;
\label{S1L}
\ea}
$\!\!$obtained from the fact that
$\tilde{P}_{2\mu}(\beta_0)$ of Eq.~(\ref{hint}) annihilates the states
$\vert\beta;N,L\rangle$ for $\beta_0=\beta$.
The mixing between the coexisting spherical and deformed $L=0$ states, 
$\vert\phi_1\rangle\equiv \vert s^N\rangle$ and 
$\vert\phi_2\rangle\equiv \vert \beta;N,L=0\rangle$, 
can be studied by first transforming to an orthonormal basis 
\ba
\vert \Psi_1\rangle &=& \vert \phi_1\rangle 
\;\; , \;
\vert \Psi_2\rangle = 
(1-r_{12}^2)^{-1/2}\,
\Bigl (\vert\phi_2\rangle  - r_{12}\,\vert \phi_1 \rangle\Bigr )~,
\nonumber\\
r_{12} &=& \langle \phi_1 \vert \phi_2\rangle = 
[N!\,\Gamma^{(L=0)}_{N}(\beta)]^{-1/2} ~,
\ea
and then examining the $2\times 2$ potential energy matrix, 
\ba
K_{ij}(\beta) = 
\langle \Psi_{i}\vert\, H_{cri}\,\vert \Psi_{j} \rangle 
= E_0\,\delta_{ij} +\tilde{K}_{ij} ~,
\label{Kij}
\ea 
which reads 
\ba
\tilde{K}_{11} &=& 0\; , \; 
\tilde{K}_{12} = -c_6\,\beta^2 \bar{N}(1-r_{12}^2)^{-1/2}r_{12} ~,
\nonumber\\
\tilde{K}_{22} &=& 
\left [\tilde{E}^{(N)}_{L=0}(\beta) 
+ 2c_6\,\beta^2\bar{N}r_{12}^2\right ]
(1-r_{12}^2)^{-1}~,
\quad
\label{tilKij}
\ea 
with $\bar{N}=N(N-1)$. 
The derived eigenvalues of the matrix 
serve as eigenpotentials, $E^{(\pm)}_{L=0}(\beta)$, and the corresponding 
eigenvectors, $\vert \Phi^{(\pm)}_{L=0}\rangle$, 
are identified with the ground ($0^{+}_1$) and excited ($0^{+}_{i}$) 
$L=0$ states. 
The deformed states $\vert \beta;N,L\rangle$ of Eq.~(\ref{wfqpt1}) 
with $L>0$ are identified 
with excited members of the ground-band ($L^{+}_1$) with energies given by 
$E_{L}^{(N)}(\beta)$, Eq.~(\ref{eneL}).
\begin{table*}
\centering
\caption{
Excitation energies (in units of $E(2^{+}_1)=1$) and B(E2) values 
[in units of $B(E2; 2^{+}_{1}\to 0^{+}_1=1)$] 
for selected terms in the critical Hamiltonian, 
Eqs.~(\ref{resol}),(\ref{hint}),(\ref{hcol}), 
with (i)~$c_3/h_2=0.05$, (ii)~$c_5/h_2 = 0.1$, (iii)~$c_6/h_2=0.05$, 
and $N=10$, $\beta_0=1.3$, $E2$ parameter $\chi = -\sqrt{7}/2$. 
The entries in square brackets [$\ldots$] are estimates based on the 
$L$-projected states, Eq.~(\ref{wfqpt1}), 
with (i)~$\beta=1.327$, (ii)~$\beta=1.318$, (iii)~$\beta=1.294$, 
determined by the global minimum of 
the respective lowest eigenvalue of the potential matrix, 
Eqs.~(\ref{Kij})-(\ref{tilKij}). 
The rigid-rotor and X(5)~{\protect\cite{iac01}} parameter-free predictions 
are shown for comparison.
\normalsize}
\vskip 10pt
\begin{tabular}{lccccc}
\hline\hline\noalign{\smallskip}
                & $c_3/h_2=0.05$ & $c_5/h_2 = 0.1$ & $c_6/h_2=0.05$ & 
rotor & X(5) \\
\noalign{\smallskip}\hline\noalign{\smallskip}
$E(4^{+}_{1})$  & 
           3.32  [3.32]  & 3.28 [3.28]   & 2.81 [2.87]    & 3.33  & 2.91 \\
$E(6^{+}_{1})$  & 
           6.98  [6.97]  & 6.74 [6.76]   & 5.43 [5.63]    & 7.00  & 5.45 \\
$E(8^{+}_{1})$  & 
          11.95 [11.95]  & 11.23 [11.29] & 8.66 [9.04]    & 12.00 & 8.51 \\
$E(10^{+}_{1})$ & 
          18.26 [18.26]  & 16.58 [16.69] & 12.23 [12.83]  & 18.33 & 12.07 \\
$E(0^{+}_{2})$  & 
           6.31  [6.30]  & 6.01 [5.93]   & 4.56 [5.03]    &       & 5.67 \\
\noalign{\smallskip}\hline\noalign{\smallskip}
$B(E2; 4^{+}_{1}\to 2^{+}_{1})$ &
           1.40 [1.40]   & 1.40 [1.40]  & 1.46 [1.45]  & 1.43  & 1.58 \\
$B(E2; 6^{+}_{1}\to 4^{+}_{1})$ & 
           1.48 [1.48]   & 1.48 [1.48]  & 1.55 [1.53]  & 1.57  & 1.98 \\
$B(E2; 8^{+}_{1}\to 6^{+}_{1})$ & 
           1.45 [1.45]   & 1.45 [1.45]   & 1.53 [1.51]  & 1.65  & 2.27 \\
$B(E2; 10^{+}_{1}\to 8^{+}_{1})$  & 
           1.37 [1.37]   & 1.37 [1.37]  & 1.44 [1.42]  & 1.69  & 2.61 \\
$B(E2; 0^{+}_{2}\to 2^{+}_{1})$ & 
           0.003 [0.003] & 0.003 [0.004] & 0.24 [0.18] &       & 0.63 \\
\noalign{\smallskip}\hline\hline
\end{tabular}
\end{table*}
E2 matrix elements between these states can be evaluated in closed form.
For example, for the $E2$ operator, 
$T(E2)=d^{\dagger}s+s^{\dagger}\tilde{d} 
+ \chi(d^{\dagger}\tilde{d})^{(2)}$, 
the necessary matrix elements for transitions involving the 
$L=0^{+}_1,2^{+}_1,0^{+}_{i}$ states are
$T_1\equiv\langle \beta; N, L^{\prime}=2\vert\vert\, T(E2)\, 
\vert \vert \beta; N,L=0\rangle$ and
$T_2\equiv\langle \beta; N, L^{\prime}=2\vert\vert\, T(E2)\, 
\vert\vert s^N \rangle$, 
\ba
T_1 &=&
\frac{\beta
[ \Gamma^{(0)}_{N-1}(\beta) 
+ \left ( 1- \beta\,\bar{\chi}\right)\Gamma^{(2)}_{N-1}(\beta) ]}
{\left [ \Gamma^{(2)}_{N}(\beta)\,\Gamma^{(0)}_{N}(\beta) \right ]^{1/2}}~,
\nonumber\\
T_2 &=& 
\beta N/
[ N!\,\Gamma^{(2)}_{N}(\beta)\,]^{1/2}~,
\label{t1t2}
\ea
with $\bar{\chi}=\sqrt{2/7}\chi$.
The parameter 
$\beta$ in the indicated wave functions, energies, and $E2$ 
matrix elements, plays the role of an effective deformation 
whose value is chosen at the global minimum of the lowest 
eigenvalue, $E^{(-)}_{L=0}(\beta)$, of the matrix $K_{ij}(\beta)$ 
(\ref{Kij}). As is evident from Table~I, 
this procedure leads to accurate finite-N estimates of observables 
at the critical-point. 
The rotational terms in $H_c$ (\ref{hcol}) 
affect the value of the effective deformation. This highlights the importance 
of the coupling between the order-parameter ($\beta)$ 
and the soft- (rotational) modes at the 
critical-point of a quantum phase transition~\cite{belitz05}. 
The characteristic spectra, discussed above, 
of these rotational terms can now be understood from 
their contribution to 
$E^{(N)}_{L}(\beta)$, Eq.~(\ref{eneL}), and 
$K_{ij}(\beta)$, Eqs.~(\ref{Kij})-(\ref{tilKij}). 
As observed from Eq.~(\ref{S1L}), 
$S^{(N)}_{1,L}$ shows an 
$L(L+1)$ behaviour near the $SU(3)$ value $\beta=\sqrt{2}$ 
and, consequently, 
the contribution of the $c_3$-term to $E^{(N)}_{L}(\beta)$ 
is linear in $L(L+1)$ while that of the 
$c_5$- and $c_6$- terms is parabolic. 
The rigid- (non-rigid) rotor-like spectrum 
of the $c_3$- ($c_5$-) terms then follows from the fact that 
they contribute only to the diagonal matrix-element 
$K_{22}(\beta)$ in Eq.~(\ref{tilKij}). 
The $c_6$-term contributes both to diagonal and non-diagonal 
matrix elements, thus controls the mixing which is essential for 
obtaining an $X(5)$-like spectrum. 

To summarize, we have considered properties of a first-order quantum 
shape-phase transition in a finite system from an algebraic perspective. 
A suitable critical-point Hamiltonian was 
constructed by resolving it into intrinsic and collective parts. The 
intrinsic part generates an energy surface with two degenerate 
minima separated by an arbitrary barrier, and its spectrum exhibits 
coexistence of spherical and deformed states. The collective part contains 
kinetic rotational terms which affect the 
rotational splittings and mixing. 
Spectral signatures for the case of a high-barrier have been 
shown to differ from those of a low-barrier. 
The dynamics at the critical-point can be described by 
an effective deformation determined by variation after 
projection, combined with two-level mixing of $L=0$ states. 
Wave functions of 
a particular analytic form can be used to derive estimates for energies 
and quadrupole rates at the critical-point. 
The intrinsic-collective resolution 
constitute an efficient method for 
studying shape-phase transitions, since the derived 
Hamiltonian is tailored to reproduce a given energy surface 
which, in-turn, governs the nature of the phase transition. 
Although we have treated explicitly the IBM with one type of bosons, 
the tools developed are applicable to other finite systems 
described by similar algebraic models, 
{\it e.g.}, the proton-neutron version of the IBM for nuclei~\cite{ibm} 
and the vibron model for molecules~\cite{vibron}. 
This work was supported by the Israel Science Foundation.

\end{document}